\documentclass[12pt]{article}

\usepackage[english]{babel}
\usepackage[cp1251]{inputenc}
\usepackage[T1]{fontenc}

\pdfoutput=1
\usepackage{makeidx}
\usepackage{amssymb}
\usepackage{amsfonts}
\usepackage{amsmath}
\usepackage{graphicx}
\usepackage{setspace}
\usepackage{authblk}
\usepackage{units}
\usepackage{appendix}
\usepackage{xparse}
\usepackage{cite}
\usepackage{hyperref}
\usepackage[left=2.1cm,top=2.5cm,right=2.1cm]{geometry}

\begin{document}

\title{Coherent states of finite-level systems}
\author{A. I. Breev$^{1}$\thanks{
		breev@mail.tsu.ru}, D. M. Gitman$^{2,3}$\thanks{
		dmitrygitman@hotmail.com}
	, \\
	$^{1}$\small{Department of Physics, Tomsk State University, \\
		Lenin ave. 36, 634050 Tomsk, Russia;}\\
	$^{2}$ P.N. Lebedev Physical Institute, \\
	53 Leninskiy ave., 119991 Moscow, Russia.\\
	$^{3}$ Institute of Physics, University of S\~{a}o Paulo, \\
	Rua do Mat\~{a}o, 1371, CEP 05508-090, S\~{a}o Paulo, SP, Brazil.}

\maketitle

\begin{abstract}
	A method for constructing coherent states (\textrm{CS}) of finite-level systems with a given angular momentum is proposed. To this end we generalize the known spin equation (\textrm{SE}) to an infinite-dimensional Fock space. The equation describes a special quadratic system in the latter space. Its projections on $d$-dimensional subspaces, represent analogs of \textrm{SE} for $d$-dimensional systems in an external electromagnetic field which describe $d$-dimensional systems with a given angular moment. Using a modification of Malkin-Manko method developed in our earlier work, we construct the corresponding \textrm{CS} for the total quadratic system. Projections of the later \textrm{CS} on finite-dimensional subspaces we call angular moment \textrm{CS} (\textrm{AMCS}) of finite-level systems. The \textrm{AMCS} have a clear physical meaning, they obey the Schr\"odinger for a $d$-dimensional system with a given angular moment $j=\left(d-1\right)/2$ in an external electromagnetic field. Their possible exact solutions are constructed via exact solutions of the \textrm{SE} in $2$-dimensional space. The latter solutions can be found analytically and are completely described in our earlier works. A one subset of \textrm{AMCS} can be related to Perelomov spinning \textrm{CS} (\textrm{PSCS}). This reflects the fact that the set of possible \textrm{AMCS} is wider than the set of \textrm{PSCS}. \textrm{AMCS} states in a constant magnetic field are constructed. Some of them coincide with the Bloch \textrm{CS}.	
	
	
\end{abstract}

\section{Introduction\label{S1}}

Coherent states (\textrm{CS}) play an important role in modern quantum theory as states
that provide a natural relation between quantum mechanical and classical or
semi-classical descriptions. They have a number of useful properties and, as a
consequence, a wide range of applications, e.g., in radiation theory, in
quantization theory, in condensed matter physics, in quantum computations, and
so on, see, e.g., Refs. \cite{KlaSu68,MalMa68,Gazea}.

In the beginning \textrm{CS} were constructed for nonrelativistic quantum systems with
describe spectrum basically oscillator like or quadratic systems in
infinite-dimensional Hilbert spaces \cite{Schro26,Rober30,Neuma31,Glaub63}.
Such \textrm{CS} turned out to be orbits of the Heisenberg-Weyl group. That
observation allowed one to formulate by analogy some general definition of
\textrm{CS} for any Lie group \cite{Perel72,Raset75,Perel86} as orbits of
the group factorized with respect to a stationary subgroup. In particular
\textrm{CS} of the $SU(N)$ and $SU(N,1)$ group were constructed in Refs. \cite{Gilmo72,GitSh93}.

In the present article, we consider the problem of constructing \textrm{CS} of
finite-level systems. Finite-level systems have always played an important
role in quantum physics. For example, in the semiclassical theory of laser
beams, in optical resonance, in problems of interaction of an assembly of
two-level atoms with a transverse electromagnetic field, and so on
\cite{Rabi36,Allen75}. $2$- and $4$-level systems have attract even more
attention, due to their relationship to problems of quantum computations, see
e.g. Refs. \cite{Nielson 2000,BagBaGL07,BalGi08,BalGaG09,BagBaGL 11}. In
particular, many problems in quantum physics which can be dealt with in terms
of the $2$-level systems were studied by many authors using different methods,
see, for example, \cite{BagBaGW01,BagGiBL05}. In this problem, the computation
is performed by the manipulation of the so-called one and $2$-qubit gates
\cite{Bremner 2002}.

In the description of an assembly of two-level atoms, atomic coherent states
are defined which have properties analogous to those of the field coherent
states \cite{Arecchi 1972}. In fact, to each property of the atomic coherent
states there exists a corresponding property of the field coherent states. The
Dicke states \cite{Dicke 1954}, that have been used in the study of
superradiance, they are a close relationship to the Fock states of the
free-field problem and by rotating this states through an angle $(\theta,\phi)$ one obtain the atomic coherent states (or Block states \cite{Bloch 1946}).

The article is organized as follows. In Secs. \ref{S2.1} and \ref{S2.2} we
discuss states of $2$- and $d$-dimensional systems with the help of angular
momentum operators that are acting in an infinite dimensional Fock space. In
Sec. \ref{S2.3} we construct a generating spin equation (\textrm{GSE}) in
the infinite dimensional Fock space generalizing a spin equation
(\textrm{SE}) for a $2$-dimensional system in an external electromagnetic
field. By projecting the \textrm{GSE} on $d$-dimensional subspaces, we obtain an analog
of \textrm{SE} for $d$-dimensional systems in an external electromagnetic
field. The Hamiltonian of \textrm{GSE} turns out to be quadratic in some Bose
creation and annihilation operators. This fact allows us to construct in Sec.
\ref{S3} generalized \textrm{CS} using Malkin-Manko integral of motion method
\cite{MM} and its development presented in Refs. \cite{BG}. Then, we study
projections of the constructed \textrm{CS} on the $d$-dimensional subspaces.
We demonstrate that a subset of these projections can be related to know spin
\textrm{CS} of $SU(2)$ group derived by Perelomov, see e.g. Ref.
\cite{Perel72}. The important case of $2$-level system in a constant magnetic
field is considered in detail in Sec. \ref{S4.1}. In the last Sect. \ref{S4}, we summarize and discuss the main results. Some auxiliary mathematical
formulas are placed in Appendices.

\section{Finite-level systems\label{S2}}

\subsection{2-level systems. Spin equation\label{S2.1}}

We recall that states of a $2$-level quantum system are two columns
$\left\vert \mathbf{\Psi}\right\rangle^{(2)}$,
\begin{equation}
	\left\vert \mathbf{\Psi}\right\rangle ^{\left(  2\right)  }=\left(
	\begin{array}
		[c]{c}%
		\psi_{1}\\
		\psi_{2}%
	\end{array}
	\right)  \ . 
	\label{1.0}
\end{equation}
The Hilbert space describing such a system is a complex $2$-dimensional space
$\mathcal{R}^{(2)}=\mathbb{C}^{2}$. As well known physical
examples of $2$-level systems are spin-$1/2$ particle, and polarization 
states of transversal photons.

A scalar product of two vectors $\left\vert \mathbf{\Psi}^{\prime
}\right\rangle ^{(2)}$ and $\left\vert \mathbf{\Psi
}\right\rangle ^{(2)}$ in the space $\mathcal{R}^{(2)}$ reads: 
$^{(2)}\langle\mathbf{\Psi}^{\prime}\left\vert \mathbf{\Psi}\right\rangle^{(2)}=\psi_{1}^{\prime\ast}\psi_{1}+\psi_{2}^{\prime\ast}\psi_{2}$. An orthogonal basis $\left\vert a\right\rangle$, ($a=1$, $2$) in $\mathcal{R}^{(2)}$ can be chosen as:
\begin{equation}
	\left\vert 1\right\rangle =\left(
	\begin{array}
		[c]{c}%
		1\\
		0
	\end{array}
	\right)  ,\ \left\vert 2\right\rangle =\left(
	\begin{array}
		[c]{c}
		0\\
		1
	\end{array}
	\right)  ,\ \langle a\left\vert a^{\prime}\right\rangle =\delta_{aa^{\prime}%
	},\ \sum_{a=1}^{2}\left\vert a\right\rangle \langle a|\ =I_{2\times2}\ ,
	\label{1.0b}
\end{equation}
where $I_{2\times2}$ is $2\times2$ unit matrix.

A quantum dynamic of a $2$-level system in a time-dependent background, given
by a real $3$-dimensional vector $\mathbf{F}(t)  =\left(
F_{1}(t)  ,F_{2}(t)  ,F_{3}(t)\right)$, is described by the Schr\"odinger equation of the form\footnote{Throughout the text $\hbar=c=1$, Latin indices are
	$i,j,k,...=1,2,3$ and Greek indices are $\alpha,\beta,\eta,...=0,1$. Unless
	otherwise indicated, summation over repeated indices is implied.}:
\begin{equation}
	i\partial_{t}\left\vert \mathbf{\Psi}\left(  t\right)  \right\rangle ^{\left(
		2\right)  }=H^{\left(  2\right)  }(t)\left\vert \mathbf{\Psi}\left(  t\right)
	\right\rangle ^{\left(  2\right)  },\text{ \ }H^{\left(  2\right)
	}(t)=\mathbf{sF}\left(  t\right)  \ , 
	\label{1.1}
\end{equation}
were $\mathbf{s}=\left(s_{1},s_{2},s_{3}\right)$ is the $1/2$ angular momentum operator,
\begin{equation}
	\left[  s_{i},s_{j}\right]  =i\varepsilon_{ijk}s_{k},\ \mathbf{s}%
	=\frac{\boldsymbol{\sigma}}{2},\ \boldsymbol{\sigma}=\left(  \sigma_{1}%
	,\sigma_{2},\sigma_{3}\right)  \ , 
	\label{1.2}
\end{equation}
where $\sigma_{i}$ are Pauli matrices\footnote{
	\begin{equation}
		\sigma_{1}=\left(
		\begin{array}
			[c]{cc}
			0 & 1\\
			1 & 0
		\end{array}
		\right)  ,\text{ \ }\sigma_{2}=\left(
		\begin{array}
			[c]{cc}
			0 & -i\\
			i & 0
		\end{array}
		\right)  ,\text{ \ }\sigma_{3}=\left(
		\begin{array}
			[c]{cc}
			1 & 0\\
			0 & -1
		\end{array}
		\right)  ,
	\end{equation}
} and $\varepsilon_{ijk}$ is the fully antisymmetric tensor with the
normalization $\varepsilon_{123}=1$. The operators $s_{i}$ implement a
representation of the Lie algebra $su(2)$ in the space $\mathcal{R}^{(2)}$. Thus, the Hamiltonian $H^{(2)}(t)$ is given by the $2\times 2$ Hermitian matrix:
\begin{equation}
	H_{\alpha\beta}^{\left(  2\right)  }(t)=\frac{1}{2}\left(
	\begin{array}
		[c]{cc}%
		F_{3}\left(  t\right)  & F_{1}\left(  t\right)  -iF_{2}\left(  t\right) \\
		F_{1}\left(  t\right)  +iF_{2}\left(  t\right)  & -F_{3}\left(  t\right)
	\end{array}
	\right)  ,\ H^{\left(  2\right)  }(t)=H^{\left(  2\right)  }(t)^{\dagger}\ .
	\label{1.4}
\end{equation}

In particular, for a frozen in the space nonrelativistic electron of the mass
$m_{e}$ and of the charge $-e$ ($e>0$), (and spin $1/2$) interacting with an
external magnetic field 
$\mathbf{B}(t)  =\left(  B_{1}(t)  ,B_{2}(t)  ,B_{3}(t)  \right)$, the
vector $\mathbf{F}(t)$ and the corresponding Hamiltonian $H^{(2)}(t)$ reads:
\begin{equation}
	\mathbf{F}\left(  t\right)  =\frac{e}{m_{e}}\mathbf{B}\left(  t\right)
	,\ H^{\left(  2\right)  }(t)\mathcal{=}\frac{e}{m_{e}}\mathbf{sB}(t)
	\ . 
	\label{1.5}
\end{equation}
For this reason, equation (\ref{1.1}) is often called the \textrm{SE}. Its
possible exact solutions were studied in detail in Ref. \cite{BagGiBL05}.

\subsection{d-level systems\label{S2.2}}

States of a $d$-level quantum system are $d$-component columns $\left\vert
\mathbf{\Psi}\right\rangle^{(d)}$,
\begin{equation}
	\left\vert \mathbf{\Psi}\right\rangle ^{\left(  d\right)  }=\left(
	\begin{array}
		[c]{cccc}%
		\psi_{1}^{(d)} & \psi_{2}^{(d)} & \cdots & \psi_{d}^{(d)}
	\end{array}
	\right)  ^{T},\ d\in\mathbb{N\ }. 
	\label{1.5b}
\end{equation}
The Hilbert space describing such a system is the complex $d$-dimensional
space $\mathcal{R}^{(d)}=\mathbb{C}^{d}$.

Below, we consider a specific basis in $d$-dimensional space 
$\mathcal{R}^{(d)}$ and a possibility to construct an analog of \textrm{SE}  (\ref{1.1}) for $d$-level systems. To this end, it is convenient to
consider an infinite-dimensional Fock space $\mathcal{R}^{(\infty)}$ (with vectors $\left\vert \Psi\right\rangle \in \mathcal{R}^{\left(  \infty\right)}$) generated by two kinds of Bose annihilation and creation operators, $\hat{a}_{\alpha}$ and $\hat{a}_{\alpha}^{\dagger}$, $\alpha=1,2$,
\begin{equation}
	\left[  \hat{a}_{\alpha},\hat{a}_{\beta}\right]  =\left[  \hat{a}_{\alpha
	}^{\dagger},\hat{a}_{\beta}^{\dagger}\right]  =0,\text{ \ }\left[  \hat
	{a}_{\alpha},\hat{a}_{\beta}^{\dagger}\right]  =\delta_{\alpha\beta}%
	,\ \alpha,\beta=1,2\ . 
	\label{1.6}
\end{equation}
An occupation number basis $\left\vert n_{1},n_{2}\right\rangle$, $n_{1},n_{2}\in\mathbb{N}$, in the space $\mathcal{R}^{(\infty)}$, is defined as:
\begin{align}
	&  \left\vert n_{1},n_{2}\right\rangle =\frac{\left(  \hat{a}_{1}^{\dagger
		}\right)  ^{n_{1}}\left(  \hat{a}_{2}^{\dagger}\right)  ^{n_{2}}}{\sqrt
		{n_{1}!n_{2}!}}\left\vert 0,0\right\rangle ,\text{ \ }\hat{a}_{\alpha
	}\left\vert 0,0\right\rangle \ =0,\ \alpha=1,2\ ,\nonumber\\
	&  \hat{a}_{1}\left\vert n_{1},n_{2}\right\rangle =\sqrt{n_{1}}\left\vert
	n_{1}-1,n_{2}\right\rangle ,\ \ \hat{a}_{2}\left\vert n_{1},n_{2}\right\rangle
	=\sqrt{n_{2}}\left\vert n_{1},n_{2}-1\right\rangle \ ,\nonumber\\
	&  \hat{a}_{1}^{\dagger}\left\vert n_{1},n_{2}\right\rangle =\sqrt{n_{1}%
		+1}\left\vert n_{1}+1,n_{2}\right\rangle ,\ \ \hat{a}_{2}^{\dagger}\left\vert
	n_{1},n_{2}\right\rangle =\sqrt{n_{2}+1}\left\vert n_{1},n_{2}+1\right\rangle
	\ , 
	\label{1.7}
\end{align}
and%
\begin{align}
	&  \hat{n}_{\alpha}\left\vert n_{1},n_{2}\right\rangle =n_{\alpha}\left\vert
	n_{1},n_{2}\right\rangle ,\text{ }\hat{n}_{\alpha}=\hat{a}_{\alpha}^{\dagger
	}\hat{a}_{\alpha},\ \alpha=1,2\ ;\nonumber\\
	&  \left\langle m_{2},m_{1}|n_{1},n_{2}\right\rangle =\delta_{m_{1}n_{1}%
	}\delta_{m_{2}n_{2}},
	{\displaystyle\sum\limits_{n_{1},n_{2}=0}^{\infty}}
	\left\vert n_{1},n_{2}\right\rangle \left\langle n_{2},n_{1}\right\vert =1\ ,
	\label{1.7b}
\end{align}
where $\left\vert 0,0\right\rangle$ is the vacuum vector in the space
$\mathcal{R}^{(\infty)}$.

One can introduce angular momentum operators $\mathbf{\hat{S}}=\left(\hat
{S}_{1},\hat{S}_{2},\hat{S}_{3}\right)$ acting in the space $\mathcal{R}^{(\infty)}$,
\begin{equation}
	\mathbf{\hat{S}}=\hat{a}_{\alpha}^{\dagger}\mathbf{s}^{\alpha\beta}\hat
	{a}_{\beta},\text{ \ }\hat{S}_{i}=\frac{1}{2}\hat{a}_{\alpha}^{\dagger}%
	\sigma_{i}^{\alpha\beta}\hat{a}_{\beta}\ , 
	\label{1.8}
\end{equation}
where $\mathbf{s}$ are given by Eq. (\ref{1.2}), see e.g. \cite{Schwi65}.
These operators satisfy the following relations:
\begin{align}
	&  \left[  \hat{S}_{i},\hat{S}_{j}\right]  =i\varepsilon_{ijk}\hat{S}%
	_{k},\text{ }\left[  \hat{S}_{3},\hat{S}_{\pm}\right]  =\pm\hat{S}_{\pm
	},\text{ \ }\left[  \hat{S}_{+},\hat{S}_{-}\right]  =2\hat{S}_{3},\text{
		\ }\left[  \mathbf{\hat{S}}^{2},\hat{S}_{3}\right]  =0\ ,\nonumber\\
	&  \hat{S}_{+}=\hat{S}_{1}+i\hat{S}_{2}=\hat{a}_{1}^{\dagger}\hat{a}%
	_{2},\text{ \ }\hat{S}_{-}=\hat{S}_{1}-i\hat{S}_{2}=\hat{a}_{2}^{\dagger}%
	\hat{a}_{1}\ ,\nonumber\\
	&  \hat{S}_{3}=\frac{1}{2}\left(  \hat{S}_{+}\hat{S}_{-}-\hat{S}_{-}\hat
	{S}_{+}\right)  =\frac{1}{2}\left(  \hat{n}_{1}-\hat{n}_{2}\right)  ,\text{
		\ }\mathbf{\hat{S}}^{2}=\frac{\hat{N}}{2}\left(  \frac{\hat{N}}{2}+1\right)
	,\ \hat{N}=\sum_{\alpha}\hat{a}_{\alpha}^{\dagger}\hat{a}_{\alpha}=\hat{n}%
	_{1}+\hat{n}_{2}\ ,\nonumber\\
	&  \hat{S}_{+}\left\vert n_{1},n_{2}\right\rangle =\sqrt{n_{2}\left(
		n_{1}+1\right)  }\left\vert n_{1}+1,n_{2}-1\right\rangle\,,\nonumber\\
	& \hat{S}_{-}\left\vert n_{1},n_{2}\right\rangle =\sqrt{n_{1}\left(  n_{2}+1\right)
	}\left\vert n_{1}-1,n_{2}+1\right\rangle \ . 
	\label{1.9}
\end{align}

Eigenvectors of the commuting operators $\mathbf{\hat{S}}^{2}$ and $\hat
{S}_{3}$ we denote as $\overline{\left\vert s,m\right\rangle}$,
\begin{align}
	&  \mathbf{\hat{S}}^{2}\overline{\left\vert s,m\right\rangle }\ =s\left(
	s+1\right)  \overline{\left\vert s,m\right\rangle },\text{ \ }%
	s=0,1/2,1,3/2,\cdots,\text{\ }\ \hat{N}\overline{\left\vert s,m\right\rangle
	}\ =2s\overline{\left\vert s,m\right\rangle }\ ,\nonumber\\
	&  \hat{S}_{3}\overline{\left\vert s,m\right\rangle }\ =m\overline{\left\vert
		s,m\right\rangle },\text{ }m=-s,-s+1,\cdots,s\ ,\nonumber\\
	&  \hat{S}_{+}\ \overline{\left\vert s,m\right\rangle }=\sqrt{\left(
		s-m\right)  \left(  s+m+1\right)  }\overline{\left\vert s,m\right\rangle
	},\ \ \hat{S}_{-}\overline{\left\vert s,m\right\rangle }\ =\sqrt{\left(
		s+m\right)  \left(  s-m+1\right)  }\overline{\left\vert s,m\right\rangle }\ .
	\label{1.10}
\end{align}
There is one-to-one correspondence between the vectors 
$\overline{\left\vert s,m\right\rangle}$ and the occupation number basis 
$\left\vert n_{1}, n_{2}\right\rangle$,
\begin{align}
	&  \overline{\left\vert s,m\right\rangle }\ =\left\vert s+m,s-m\right\rangle
	,\text{ \ }\left\vert n_{1},n_{2}\right\rangle =\overline{\left\vert
		\frac{n_{1}+n_{2}}{2},\frac{n_{1}-n_{2}}{2}\right\rangle }\ ,\nonumber\\
	&  \overline{\left\vert 0,0\right\rangle }=\left\vert 0,0\right\rangle \ .
	\label{1.11}
\end{align}

The space $\mathcal{R}^{(\infty)}$ can be represented as a direct orthogonal sum\footnote{Let H be a Hilbert space, and let $H_1$ and $H_2$ be its orthogonal closed subspaces, $H_1\perp H_2$. If any vector $\xi\in H$ can be uniquely represented as a
	sum of two vectors $\xi_1 + \xi_2$, where $\xi_1\in H_1$, and $\xi_2\in H_2$, then
	$H=H_{1}\oplus H_{2}$.} of the $d$-dimensional spaces $\mathcal{R}^{(d)}$,
\begin{equation}
	\mathcal{R}=\bigoplus_{d=1}^{\infty}\mathcal{R}^{\left(  d\right)  }\ .
	\label{1.12}
\end{equation}
Thus, any vector $\left\vert \Psi\right\rangle \in\mathcal{R}^{(\infty)}$ can be represented as:
\begin{equation}
	\left\vert \Psi\right\rangle =\sum_{d=1}^{\infty}\left\vert \Psi\right\rangle
	^{\left(  d\right)  },\ \ \left\vert \Psi\right\rangle ^{\left(  d\right)
	}=P^{(d)}\left\vert \Psi\right\rangle =\sum_{m=-\frac{d-1}{2}}^{m=\frac
		{d-1}{2}}\psi_{\frac{d+1}{2}-m}\overline{\left\vert \frac{d-1}{2}%
		,m\right\rangle }\ \ \in\mathcal{R}^{\left(  d\right)  }\ , 
	\label{1.13}
\end{equation}
where $\psi_{k}$ are some decomposition coefficients. By $P^{(d)}$ in Eq.
(\ref{1.13}) we denote the operator of projection of vectors from
$\mathcal{R}^{\left(  \infty\right)  }$ onto the $d$-dimensional subspace
$\mathcal{R}^{\left(  d\right)  }$. All the states that belong to
$\mathcal{R}^{\left(  d\right)  }$ are eigenvectors of the operator 
$\hat{d}=\hat{N}+1$,
\begin{equation}
	\hat{d}\left\vert \Psi\right\rangle ^{\left(  d\right)  }=d\left\vert
	\Psi\right\rangle ^{\left(  d\right)  },\ \ \hat{d}=\hat{N}+1\ . 
	\label{1.13b}
\end{equation}

It is convenient to introduce $d$-vectors $\overline{\left\vert a\right\rangle
}$, $a=1,\dots,d$,
\begin{equation}
	\overline{\left\vert a\right\rangle }=\overline{\left\vert \frac{d-1}{2}%
		,\frac{d-1}{2}-a+1\right\rangle }=\left\vert d-a,a-1\right\rangle \ .
	\label{1.13bb}
\end{equation}
As follows from Eq. (\ref{1.7b}) these vectors form an orthogonal basis in
$\mathcal{R}^{(d)}$,
\begin{equation}
	\overline{\langle a\left\vert a^{\prime}\right\rangle }\ =\delta_{aa^{\prime}%
	},\ \sum_{a=1}^{d}\overline{\left\vert a\right\rangle }\overline{\left\langle
		a\right\vert }\ =1\ . 
	\label{1.13cc}
\end{equation}
Basis (\ref{1.13bb}) specifies $\overline{\left\vert a\right\rangle}$-representation of vectors in $\mathcal{R}^{(d)}$ as follows:
\begin{equation}
	\left\vert \Psi\right\rangle ^{\left(  d\right)  }=\sum_{a=1}^{d}\psi
	_{a}^{(d)}\ \overline{\left\vert a\right\rangle },\ \ \left\vert
	\Psi\right\rangle ^{\left(  d\right)  }\Longrightarrow\psi_{a}^{(d)}%
	=\overline{\langle a}\left\vert \Psi\right\rangle ^{(d)}\ . 
	\label{1.13ee}
\end{equation}
The components $\psi_{a}^{(d)}$ form column (\ref{1.5b}). In the future, when
it does not lead to misunderstanding, we will omit the superscript $(d)$ and
write $\psi_{a}^{(d)}=\psi_{a}$.

\section{Angular momentum equations in spaces $\mathcal{R}^{(\infty)}$ and $\mathcal{R}^{(d)}$\label{S2.3}}

Starting from the \textrm{SE} (\ref{1.1}), defined for vectors $\left\vert
\Psi\left(t\right)\right\rangle^{(2)}$ in the
$2$-dimensional Hilbert space $\mathcal{R}^{(2)}$, we construct
the Schr\"odinger equation for vectors $\left\vert \Psi(t)
\right\rangle \in\mathcal{R}^{(\infty)}$, with the Hamiltonian $\hat{H}$,
\begin{equation}
	\hat{H}=\mathbf{\hat{S}F}\left(  t\right)  =\hat{a}_{\alpha}^{\dagger}\left(
	H^{\left(  2\right)  }\right)  ^{\alpha\beta}\hat{a}_{\beta}\ , 
	\label{1.13c}
\end{equation}
in fact, substituting the spin operators $\mathbf{s}$ in $H^{(2)}$ given by Eq. (\ref{1.1}) by the angular momentum operators $\mathbf{\hat{S}}$. In such a way we obtain an analog of the \textrm{SE} (\ref{1.1}) in the infinite-dimensional Hilbert space $\mathcal{R}^{(\infty)}$,
\begin{equation}
	i\partial_{t}\left\vert \Psi\left(  t\right)  \right\rangle =\hat{H}\left\vert
	\Psi\left(  t\right)  \right\rangle ,\ \ \left\vert \Psi\left(  t\right)
	\right\rangle \in\mathcal{R}^{\left(  \infty\right)  }\ . 
	\label{1.14}
\end{equation}
In what follows, we call equation (\ref{1.14}) the angular momentum equation
in the infinite-dimensional Hilbert space $\mathcal{R}^{(\infty)}$, or simply the generating spin equation (GSE) in $\mathcal{R}^{(\infty)}$. Obviously, that $\hat{H}$ is a quadratic in annihilation and creation operators, $\hat{a}_{\alpha}$ and $\hat{a}_{\alpha}^{\dagger}$, $\alpha=1$, $2$, defined by Eqs. (\ref{1.6}).

Solutions $\left\vert \Psi(t)\right\rangle ^{(d)}$ of Eq. (\ref{1.14}) that belong to the space $\mathcal{R}^{(d)}$, have to satisfy the equation $\hat{d}\left\vert \Psi(t)  \right\rangle^{(d)}=d\left\vert \Psi(t)\right\rangle ^{(d)}$ and the corresponding
Schr\"odinger equation, i.e.
\begin{equation}
	i\partial_{t}\left\vert \Psi\left(  t\right)  \right\rangle ^{(d)}=\hat
	{H}\left\vert \Psi\left(  t\right)  \right\rangle ^{(d)}\ . 
	\label{1.15}
\end{equation}
In what follows, we call Eq. (\ref{1.15}) the angular momentum equation in
$\mathcal{R}^{(d)}$.

Let us find an explicit form of Eq. (\ref{1.15}) in representation
(\ref{1.13ee}).

It is demonstrated in the Appendix \textrm{A} that the space $\mathcal{R}^{(d)}$ is invariant under the angular momentum operators $\mathbf{\hat{S}}$ entering Hamiltonian (\ref{1.13c}). A restriction of the operators $\mathbf{\hat{S}}$ on the subspace $\mathcal{R}^{(d)}$ is given by operators 
$\mathbf{s}^{(d)}=\left(s_{1}^{(d)},s_{2}^{(d)},s_{3}^{(d)}\right)$ of the angular momentum $j=(d-1)/2$,
\begin{equation}
	\left[  s_{i}^{(d)},s_{j}^{(d)}\right]  =i\varepsilon_{ijk}s_{k}%
	^{(d)},\ \ s_{i}^{(d)}=\left(  \hat{S}_{i}\right)  _{ab},\ i,j=1,\dots
	,d;\ \ a,b=1,\dots,d\ , 
	\label{1.16b}
\end{equation}
where explicit form of the matrices $\left(\hat{S}_{i}\right)_{ab}$ is
presented by Eq. (\ref{B.6}).

With help of Eqs. (\ref{B.5}) and (\ref{B.6}) one can find the action of the
operator $\hat{H}$ on the vector $\left\vert \Psi\right\rangle^{(d)}$,
\begin{align}
	&  \hat{H}\left\vert \Psi\right\rangle ^{\left(  d\right)  }=\ \sum
	_{a,b=1}^{d}H_{ab}^{(d)}(t)\psi_{b}(t)\overline{\left\vert a\right\rangle
	}\ ,\ \hat{d}\ \hat{H}\left\vert \Psi\right\rangle ^{\left(  d\right)
	}=d\ \hat{H}\left\vert \Psi\right\rangle ^{\left(  d\right)  }\ ,\nonumber\\
	&  H_{ab}^{(d)}(t)=\sum_{i=1}^{3}\overline{\langle a|}\hat{S}_{i}%
	\overline{|b\rangle}F_{i}(t)=\mathbf{s}^{(d)}\mathbf{F}\left(  t\right)
	,\ a,b=1,\dots,d\ . 
	\label{1,19c}
\end{align}
Then, multiplying Eq. (\ref{1.15}) from the left on the bra-vector
$\overline{\langle a|}\ =\left\langle a-1,d-a\right\vert$, we obtain angular
momentum equation in the $\overline{\left\vert a\right\rangle}$-representation,
\begin{equation}
	i\dot{\psi}_{a}^{(d)}(t)=\sum_{b=1}^{d}H_{ab}^{(d)}(t)\psi_{b}^{(d)}%
	(t),\ \psi_{a}^{(d)}(t)=\overline{\langle a}\left\vert \Psi\right\rangle
	^{(d)},\ a=1,\dots,d\ , 
	\label{1.19d}
\end{equation}
which reads in more detail as:
\begin{align}
	&  i\dot{\psi}_{1}^{(d)}(t)=\left(  d-1\right)  F_{0}(t)\psi_{1}%
	^{(d)}(t)+\sqrt{d-1}F_{(-)}(t)\psi_{2}^{(d)}(t)\ ,\label{1.53a}\\
	&  i\dot{\psi}_{a}^{(d)}(t)=\sqrt{(a-1)(d-a+1)}F_{(+)}(t)\psi_{a-1}%
	^{(d)}(t)\nonumber\\
	&+\left(  d-2a+1\right)  F_{0}(t)\psi_{a}^{(d)}(t)+\sqrt
	{a(d-a)}F_{(-)}(t)\psi_{a+1}^{(d)}(t),\ a=2,\dots,d-1\ ,\label{1.53b}\\
	&  i\dot{\psi}_{d}^{(d)}(t)=-\left(  d-1\right)  F_{0}(t)\psi_{d}%
	^{(d)}(t)+\sqrt{d-1}F_{(+)}(t)\psi_{d-1}^{(d)}(t)\ ,\label{1.53c}\\
	&  F_{0}(t)=F_{3}(t)/2,\ \ F_{(\pm)}(t)=\left[  F_{1}\left(  t\right)  \pm
	iF_{2}\left(  t\right)  \right]  /2\ .\nonumber
\end{align}

Thus, a restriction of the angular momentum equation (\ref{1.14}) in
$\mathcal{R}^{(\infty)}$ to the subspace $\mathcal{R}^{(d)}$ induced the 
Schr\"odinger equation (\ref{1.19d}) for a $d$-level
system in an external field. Eqs. (\ref{1.53a})--(\ref{1.53c}) are ordinary
differential equation in time.

The angular momentum equation (\ref{1.19d}) generalizes \textrm{SE}
(\ref{1.1}) to the case of higher dimension $d>2$. Besides, the dynamic
symmetry group of the $d$-dimensional system described by this equation is the
$SU(2)$ group (which holds also true for the \textrm{SE} (\ref{1.1})). This follows from the fact that the corresponding Hamiltonian is
linear in the generators $\mathbf{s}^{(d)}$ of the unitary irreducible
representation of the Lie algebra $su(2)$ of the group
$SU(2)$. It is important that the Hamiltonian (\ref{1,19c}) commutes
with the operator of the square of the angular momentum operator
\[
	s_{(d)}^{2}=\left[  s_{1}^{(d)}\right]  ^{2}+\left[  s_{2}^{(d)}\right]
	^{2}+\left[  s_{3}^{(d)}\right]  ^{2}\ .
\]

Let us find conditions for the functions $F_{0}(t)$ and $F_{(\pm)}(t)$ that
provide the existence of solutions $\psi_{a}^{(d)}(t)$ in each subspace
$\mathcal{R}^{(d)}$.

First, we consider the case when $F_{(-)}(t)$ is not identically zero. In this
case, the set of $d-1$\ equations (\ref{1.53a})--(\ref{1.53b}) turns out to be
identity if the functions $\psi_{2}^{(d)}(t)$,$\cdots$, $\psi_{d}^{(d)}(t)$
are expressed via the function $\psi_{1}^{(d)}(t)$ and its derivatives up to
$d-1$ order due to the following recurrent conditions:
\begin{align}
	&  \psi_{2}^{(d)}(t)=\frac{1}{F_{(-)}(t)}\left[  \frac{i\dot{\psi}_{1}%
		^{(d)}(t)}{\sqrt{d-1}}-\sqrt{d-1}F_{0}(t)\psi_{1}^{(d)}(t)\right]
	,\ \psi_{a+1}^{(d)}(t)=\frac{1}{F_{(-)}(t)}\left[  \frac{i\dot{\psi}_{a}%
		^{(d)}(t)}{\sqrt{a(d-a)}}\right. \label{1.54a}\\
	&  \left.  \frac{\left(  2a-d-1\right)  }{\sqrt{a(d-a)}}F_{0}(t)\psi_{a}%
	^{(d)}(t)-\sqrt{\frac{(a-1)(d-a+1)}{a(d-a)}}F_{(+)}(t)\psi_{a-1}%
	^{(d)}(t)\right]  ,\ a=2,\dots,d-1\ . 
	\label{1.54b}
\end{align}
Substituting Eqs. (\ref{1.54a}) and (\ref{1.54b}) into Eq. (\ref{1.53c}) we
obtain a linear ordinary $d$-order differential equation for the function
$\psi_{1}^{(d)}(t)$ with coefficients that depend on time via the functions
$F_{0}(t)$ and $F_{(\pm)}(t)$ and their derivatives up to $d-1$ order. Thus,
infinite differentiability of functions $F_{0}(t)$ and $F_{(\pm)}(t)$ in time
provides the existence of solutions $\psi_{a}^{(d)}(t)$ in each subspace
$\mathcal{R}^{(d)}$.

If $F_{(-)}(t)\equiv0$, then $F_{(+)}(t)\equiv 0$ and the set of Eqs.
(\ref{1.53a}) and (\ref{1.53c}) can be easily integrated,
\[
	\psi_{a}^{(d)}(t)=c_{a}^{(d)}e^{-i(d-2a+1)\Xi(t)},\ \ \Xi(t)=\int^{t}%
	F_{0}(t^{\prime})dt^{\prime},\ \ a=1,\dots,d\ ,\
\]
where $c_{a}^{(d)}$ are integration constant. Here it is sufficient the
existence of a primitive function $\Xi(t)$ for $F_{0}(t)$. We meet this case
considering $2$-level system placed in a constant magnetic field, see Sec.
\ref{S4.1}.

\section{CS states of finite-level systems\label{S3}}

\subsection{Annihilation and creation operators-integrals of motion\label{3.1}%
}

As was already mentioned above, the Hamiltonian $\hat{H}$ is quadratic in the
operators $\hat{a}$ (here and in what follows, we use condense notation:
$\hat{a}=(\hat{a}_{\alpha})$ and $\hat{a}^{\dagger}=(\hat{a}_{\alpha}^{\dagger})$).
The construction of \textrm{CS} of
quadratic systems in infinite-dimensional Hilbert spaces has been considered
in numerous works. In addition to the references given in the Introduction,
the most relevant for the construction considered below are the pioneering
works of Man'ko with coauthors \cite{MalMa68,Manko72,DodMa87} and the Ref.
\cite{BagGiP2015}.

In the beginning, we pass from the operators $\hat{a}$ and $\hat{a}^{\dagger}$
to new time-dependent operators $\hat{A}=(\hat{A}_{\alpha}(t))$
and $\hat{A}^{\dagger}=(\hat{A}_{\alpha}^{\dagger}(t))$ by the
help of a linear canonical transformation:
\begin{equation}
	\hat{A}=w\hat{a}+\upsilon\hat{a}^{\dagger}+\varphi,\text{ \ }\hat{A}^{\dagger
	}=\hat{a}^{\dagger}w^{\dagger}+\hat{a}\upsilon^{\dagger}+\varphi^{\dagger
	}\ ,
	\label{1.16}
\end{equation}
Here $w=(w_{\alpha\beta}(t))$, $\upsilon=(\upsilon_{\alpha\beta}(t))$ and $\varphi=(\varphi_{\alpha}(t))$ are some time dependent complex matrices and functions respectively. In the following we shall consider only linearly independent operators $\hat{A}_{\alpha}$. To this end one of the matrices $w$ or $\upsilon$ must be
nonsingular. For $\hat{A}$ and $\hat{A}^{\dagger}$ to be annihilation and
creation operators,
\begin{equation}
	\left[  \hat{A},\hat{A}^{\dagger}\right]  =I,\text{ \ }\left[  \hat{A},\hat
	{A}\right]  =\left[  \hat{A}^{\dagger},\hat{A}^{\dagger}\right]
	=0\ ,
	\label{1.17}
\end{equation}
the matrices $w$ and $\upsilon$ have to satisfy the following conditions (see
e.g. \cite{Berez66}):
\begin{align}
	&  ww^{\dagger}-\upsilon\upsilon^{\dagger}=I,\text{ \ }w^{\dagger}%
	w-\upsilon^{T}\upsilon^{\ast}=I\ ,\nonumber\\
	&  \upsilon w^{T}=w\upsilon^{T},\text{ \ }w^{\dagger}\upsilon=\upsilon
	^{T}w^{\ast}\ ,
	\label{1.19}
\end{align}
where by the upper sign $T$ transpose matrices are denoted.

For operators $\hat{A}(t)$ and $\hat{A}^{\dagger}(t)$ to be integrals of motion which is defined by Eq. (\ref{1.14}), it is sufficient their commutativity with the operator $\hat{\Pi}=i\partial_{t}-\hat{H}$,
\begin{equation}
	\left[  \hat{\Pi},\hat{A}\right]  =\left[  \hat{\Pi},\hat{A}^{\dagger}\right]
	=0,\ \hat{\Pi}=i\partial_{t}-\hat{H}\ . 
	\label{1.20}
\end{equation}
If $\hat{U}(t)$ is an evolution operator for Eq. (\ref{1.14}),
\[
	i\partial_{t}\hat{U}\left(  t\right)  =\hat{H}\hat{U}\left(  t\right)
	,\ \hat{U}\left(  0\right)  =1\ ,
\]
then the integrals of motion $\hat{A}_{\alpha}(t)$ and 
$\hat{A}_{\alpha}^{\dagger}(t)$ satisfy the following relations:
\begin{equation}
	\hat{A}_{\alpha}\left(  t\right)  =\hat{U}\left(  t\right)  \hat{A}_{\alpha
	}\left(  0\right)  \hat{U}^{-1}\left(  t\right)  ,\ \hat{A}_{\alpha}^{\dagger
	}\left(  t\right)  =\hat{U}\left(  t\right)  \hat{A}_{\alpha}^{\dagger}\left(
	0\right)  \hat{U}^{-1}\left(  t\right)  \ . 
	\label{1.21}
\end{equation}

Calculating the commutators $\left[\hat{\Pi},\hat{A}\right]$, we obtain:
\begin{align}
	&  \left[  \hat{\Pi},\hat{A}\right]  =\left(  i\dot{w}+wH^{\left(  2\right)
	}\right)  \hat{a}+\left(  i\dot{\upsilon}-\upsilon H^{\left(  2\right)  \ast
	}\right)  \hat{a}^{\dagger}+i\dot{\varphi}=0\nonumber\\
	&  \ \Longrightarrow i\dot{w}=-wH^{\left(  2\right)  },\ \ i\dot{\upsilon
	}=\upsilon H^{\left(  2\right)  \ast},\ \ \dot{\varphi}=0\ .
	\label{1.22}
\end{align}
Without loss of the generality we can set $\varphi(t)=0$.

Introducing spinors $W_{\alpha}$ and $V_{\alpha}$,
\begin{equation}
	W_{\alpha}=\left(
	\begin{array}
		[c]{c}%
		w_{\alpha1}\\
		w_{\alpha2}
	\end{array}
	\right)  ,\ V_{\alpha}=\left(
	\begin{array}
		[c]{c}%
		\upsilon_{\alpha1}\\
		\upsilon_{\alpha2}
	\end{array}
	\right)  \ ,
	\label{1.18}
\end{equation}
we can rewrite remaining equations for $w$ and $\upsilon$ as follows:
\begin{align}
	&  i\partial_{t}W_{\alpha}=-H^{\left(  2\right)  \ast}W_{\alpha}=\Gamma
	W_{\alpha},\ \ \Gamma=-\frac{1}{2}\left(
	\begin{array}
		[c]{cc}
		F_{3} & F_{(+)}\\
		F_{(-)} & -F_{3}
	\end{array}
	\right)  =\left(  \boldsymbol{\sigma\Omega}\right)  \ ,\nonumber\\
	&  i\partial_{t}V_{\alpha}=H^{\left(  2\right)  }(t)V_{\alpha}=\Gamma^{\ast
	}V_{\alpha}=\left(  \boldsymbol{\sigma\tilde{\Omega}}\right)  V_{\alpha
	}\ ,\nonumber\\
	&  \boldsymbol{\Omega}=-\frac{1}{2}\left(  F_{1},-F_{2},F_{3}\right)
	,\ \boldsymbol{\tilde{\Omega}}=-\frac{1}{2}\left(  F_{1},F_{2},F_{3}\right)
	\ .
	\label{1.24}
\end{align}

Spin Eq. (\ref{1.24}) for zero initial conditions $\upsilon(0)=0$ has zero solution $V_{\alpha}(t)=0$ for any external field (\ref{1.5}). In what follows, we consider only the case $\upsilon(t)=0$ and $\varphi(t)=0$, which is sufficient for our
purposes. Thus, integrals of motion $\hat{A}(t)$ and $\hat{A}^{\dagger}(t)$ that are used in our further constructions have the form:
\begin{equation}
	\hat{A}\left(  t\right)  =w\left(  t\right)  \hat{a},\text{ \ }\hat
	{A}^{\dagger}\left(  t\right)  =\hat{a}^{\dagger}w^{\dagger}\left(  t\right)
	\ ,
	\label{1.25}
\end{equation}
with matrices $w(t)$ that have to be found from Eqs. (\ref{1.22}) under the condition $\det w(t)\neq0$ that provides the linear independence of integrals of motion $\hat{A}_{\alpha}(t)$. We also impose the following initial condition for Eq.
(\ref{1.24}):
\begin{equation}
	w\left(  0\right)=I\Longrightarrow\hat{A}\left(  0\right)=\hat{a}\ .
	\label{1.26a}
\end{equation}
In such a case we have:
\begin{equation}
	\hat{A}_{\alpha}\left(  t\right)  =U\left(  t,0\right)  \hat{a}U^{-1}\left(
	t,0\right)  ,\ \hat{A}_{\alpha}^{\dagger}\left(  t\right)  =U\left(
	t,0\right)  \hat{a}^{\dagger}U^{-1}\left(  t,0\right)  \ .
	\label{1.27a}
\end{equation}

\subsection{Solving defining equations\label{S3.2}}

Now we define \textrm{CS} $\left\vert Z,t\right\rangle$ in $\mathcal{R}^{(\infty)}$ as solution of the Schr\"odinger equation (\ref{1.14}) that is also eigenvectors of the annihilation operators $\hat{A}_{\alpha}(t)$ given by Eq. (\ref{1.21}). Such a vector has to satisfy the following not contradictory (due to Eq. (\ref{1.22})) to each
other equations:
\begin{align}
	&  \hat{A}_{\alpha}\left(  t\right)  \left\vert Z,t\right\rangle =Z_{\alpha
	}\left\vert Z,t\right\rangle \ ,\label{2.1}\\
	&  \hat{\Pi}\left(  t\right)  \left\vert Z,t\right\rangle =0\ . 
	\label{2.2}
\end{align}

Taking into account Eq. (\ref{1.25}), and the condition $ww^{\dagger}=I$, we
can rewrite equations (\ref{2.1}) as:
\begin{equation}
	\hat{a}_{\alpha}\left\vert Z,t\right\rangle =\ \tilde{Z}_{\alpha}\left(
	t\right)  \left\vert Z,t\right\rangle ,\ \ \tilde{Z}_{\alpha}\left(  t\right)
	=\left(  w^{-1}\right)  _{\alpha\beta}Z_{\beta}=Z_{\beta}w_{\beta\alpha}%
	^{\ast}\ .
	\label{2.3}
\end{equation}
In each time instant $t$ the state $\left\vert Z,t\right\rangle$ has the form
of CS of the Heisenberg group,
\begin{align}
	\left\vert Z,t\right\rangle  &  =\exp\left\{  -\frac{|\tilde{Z}\left(
		t\right)  |^{2}}{2}\right\}  \exp\{Z_{\beta}w_{\beta\alpha}^{\ast}a_{\alpha
	}^{+}\}|0,0\rangle\ ,\nonumber\\
	\left\vert Z,0\right\rangle  &  =|Z\rangle,\ \hat{a}_{\alpha}|Z\rangle
	=Z_{\alpha}|Z\rangle,\ \langle Z,t|Z,t\rangle=1\ .
	\label{2.5a}
\end{align}
We note also that
\begin{equation}
	\ |\tilde{Z}\left(  t\right)  |^{2}=|Z_{\beta}w_{\beta1}^{\ast}|^{2}%
	+|Z_{\beta}w_{\beta2}^{\ast}|^{2}=Z_{\alpha}^{\ast}Z_{\beta}w_{\alpha\gamma
	}w_{\gamma\beta}^{\dagger}=Z_{\alpha}^{\ast}Z_{\alpha}=\left\vert Z\right\vert
	^{2}\ .\nonumber
\end{equation}
Thus state $\left\vert Z,t\right\rangle$ takes the form:
\begin{equation}
	\left\vert Z,t\right\rangle =\exp\left\{  -\frac{\left\vert Z\right\vert ^{2}%
	}{2}\right\}  \exp\{Z_{\beta}w_{\beta\alpha}^{\ast}a_{\alpha}^{+}%
	\}|0,0\rangle\ ,
	\label{b3.2}
\end{equation}
On the other hand, Eqs. (\ref{1.27a}) imply:
\begin{equation}
	\hat{a}_{\alpha}U^{-1}\left(  t,0\right)  \left\vert Z,t\right\rangle
	=Z_{\alpha}U^{-1}\left(  t,0\right)  \left\vert Z,t\right\rangle
	\Longrightarrow\left\vert Z,t\right\rangle =U\left(  t,0\right)
	|Z\rangle\ .
	\label{2.6a}
\end{equation}
Thus, the vector $\left\vert Z,t\right\rangle$ given by Eq. (\ref{2.5a}) satisfies also the Schr\"odinger equation (\ref{2.2}).

In \textrm{CS}-representation, vector (\ref{2.5a}) reads:
\begin{equation}
	\Psi_{Z}\left(  z,t\right)  =\langle z|Z,t\rangle=\exp\left\{  -\frac
	{\left\vert z\right\vert ^{2}+\left\vert Z\right\vert ^{2}}{2}+Z_{\beta
	}w_{\beta\alpha}^{\ast}z_{\alpha}\right\}  \ .
	\label{2.6}
\end{equation}

Expanding the exponential in expression (\ref{b3.2}) into a series and acting
on the vacuum vector $|0,0\rangle$ by the creation operators, we obtain:
\begin{align}
	&  \left\vert Z,t\right\rangle =\sum_{d=1}^{\infty}\left\vert Z,t\right\rangle
	^{(d)},\ \left\vert Z,t\right\rangle ^{(d)}=P^{(d)}\left\vert Z,t\right\rangle
	=\exp\left\{  -\frac{\left\vert Z\right\vert ^{2}}{2}\right\} \nonumber\\
	&  \times\sum_{a=1}^{d}\frac{\tilde{Z}_{1}^{d-a}\tilde{Z}_{2}^{a-1}}%
	{\sqrt{(a-1)!(d-a)!}}\overline{\left\vert a\right\rangle }\in\mathcal{R}%
	^{\left(  d\right)  },\ \left(  \hat{N}+1-d\right)  \left\vert
	Z,t\right\rangle ^{(d)}=0\ . 
	\label{2.8}
\end{align}
The square of the norm of the vector $\left\vert Z,t\right\rangle ^{(d)}$ in
$\mathcal{R}^{(d)}$ does not depend on time and is determined
by the modulus $|Z|=\sqrt{|Z_{1}|^{2}+|Z_{2}|^{2}}$,
\begin{equation}
	^{(d)}\left\langle Z,t|Z,t\right\rangle ^{(d)}=e^{-\left\vert Z\right\vert
		^{2}}\frac{|Z|^{d-1}}{\left(  d-1\right)  !},\ \sum_{d=1}^{\infty}%
	\ ^{(d)}\left\langle Z,t|Z,t\right\rangle ^{(d)}=1\ . 
	\label{2.11b}
\end{equation}

The action of the operators 
$\hat{A}_{\alpha}(t)=w_{\alpha\beta}(t)\hat{a}_{\beta}$ on the vectors 
$\left\vert Z,t\right\rangle^{(d)}$ can be easily find:
\begin{align}
	&  \hat{A}_{\alpha}\left(  t\right)  \left\vert Z,t\right\rangle ^{(1)}%
	=\exp\left\{  -\frac{\left\vert Z\right\vert ^{2}}{2}\right\}  w_{\alpha\beta
	}\left(  t\right)  \hat{a}_{\beta}\left\vert 0,0\right\rangle =0\ ,\nonumber\\
	&  \hat{A}_{\alpha}\left(  t\right)  \left\vert Z,t\right\rangle ^{(d)}%
	=\exp\left\{  -\frac{\left\vert Z\right\vert ^{2}}{2}\right\}  \sum_{a=1}%
	^{d}\frac{\tilde{Z}_{1}^{d-a}\tilde{Z}_{2}^{a-1}}{\sqrt{(a-1)!(d-a)!}%
	}w_{\alpha\beta}\left(  t\right)  \hat{a}_{\beta}\overline{\left\vert
		a\right\rangle }\nonumber\\
	&  =\exp\left\{  -\frac{\left\vert Z\right\vert ^{2}}{2}\right\}  \left\{
	\sum_{a=1}^{d-1}\sqrt{d-a}\frac{\tilde{Z}_{1}^{d-a}\tilde{Z}_{2}^{a-1}}%
	{\sqrt{(a-1)!(d-a)!}}w_{\alpha1}\left(  t\right)  \overline{\left\vert
		a\right\rangle }^{(d-1)}\right.  \nonumber\\
	&  +\left.  \sum_{a=1}^{d-1}\sqrt{a}\frac{\tilde{Z}_{1}^{d-a-1}\tilde{Z}%
		_{2}^{a}}{\sqrt{(a)!(d-a-1)!}}w_{\alpha2}\left(  t\right)  \overline
	{\left\vert a\right\rangle }^{(d-1)}\right\}  \nonumber\\
	&  =\tilde{Z}_{\beta}w_{\alpha\beta}\left(  t\right)  \exp\left\{
	-\frac{\left\vert Z\right\vert ^{2}}{2}\right\}  \sum_{a=1}^{d-1}\frac
	{\tilde{Z}_{1}^{d-a-1}\tilde{Z}_{2}^{a-1}}{\sqrt{(a-1)!(d-a-1)!}}%
	\overline{\left\vert a\right\rangle }^{(d-1)}\nonumber\\
	&  =Z_{\alpha}\left\vert Z,t\right\rangle ^{(d-1)},\ \ \left\vert
	Z,t\right\rangle ^{(d-1)}=P^{(d-1)}\left\vert Z,t\right\rangle
	,\ \ d>1\ .
	\label{2.10}
\end{align}
Here
\[
	\overline{\left\vert a\right\rangle }^{(d-1)}=\overline{\left\vert
	a\right\rangle }=\left\vert d-a-1,a-1\right\rangle \in\mathcal{R}^{\left(
	d-1\right)  },a=1,\dots,d-1
\]
and the relation $\tilde{Z}_{\beta}w_{\alpha\beta}=Z_{\gamma}\left(
ww^{\dagger}\right)_{\alpha\gamma}=Z_{\alpha}$ was taken into account.

Thus, although the state $\left\vert Z,t\right\rangle$ is an eigenstate for
the operator $\hat{A}_{\alpha}(t)$, see Eq. (\ref{2.1}), its
projection $\left\vert Z,t\right\rangle ^{(d)}$ on the subspace $\mathcal{R}^{(d)}$ is not. Nevertheless, we will call the states $\left\vert Z,t\right\rangle ^{(d)}$ angular momentum \textrm{CS} (\textrm{AMCS}) in the space $\mathcal{R}^{(d)}$.

\subsection{Relation to Perelomov spin coherent states\label{S3.3}}

Let us find a relation between the \textrm{AMCS} $\left\vert Z,t\right\rangle
^{(d)}$ and Perelomov spinning \textrm{CS} (\textrm{PSCS}) of the group
$SU(2)$ introduced in Ref. \cite{Perel72}.

The (PSCS) are defined with the help of the operators 
$\mathbf{s}^{(d)}=\left(s_{1}^{(d)},s_{2}^{(d)},s_{3}^{(d)}\right)$ of the angular
momentum $j=(d-1)/2$ in the following way. First, one chooses the fixed vector
$\left\vert j,-j\right\rangle ^{(d)}=\overline{\left\vert d\right\rangle}$
that minimizes the Casimir dispersion (\ref{B.7}),
\begin{align}
	&  \left\langle j,-j|\Delta s^{(d)^{2}}|j,-j\right\rangle =j(j+1)-j^{2}%
	=j\ ,\nonumber\\
	&  s^{(d)^{2}}\left\vert j,-j\right\rangle ^{(d)}=j(j+1)\left\vert
	j,-j\right\rangle ^{(d)}\ ,\nonumber\\
	&  s_{3}^{(d)}\left\vert j,-j\right\rangle ^{(d)}=-j\left\vert
	j,-j\right\rangle ^{(d)}\ . 
	\label{3.3.1}
\end{align}

Parametrizing elements of the group $SU(2)$ by Euler angels,
$g=\left(\phi,\theta,\psi\right)$, $0\leq\phi<2\pi$, $0\leq\theta<\pi$,
$0\leq\psi<2\pi$, we consider the subgroup $U(1)$ of diagonal
matrices,
\begin{equation}
	U(1)=\left\{
	\begin{pmatrix}
		e^{i\psi/2} & 0\\
		0 & e^{-i\psi/2}%
	\end{pmatrix}
	\right\}  \ . 
	\label{3.3.2}
\end{equation}

The homogeneous space $X=SU(2)/U(1)$ is isomorphic to the
$2$-dimensional unit sphere, that is, to the set of unit vectors 
$\mathbf{n}=\left(\sin\theta\cos\phi,\sin\theta\sin\phi,\cos\theta\right)$. Using the
stereographic projection of a given sphere onto the complex plane, we will
parametrize the points of the homogeneous space $X$ using a complex number
$\zeta=-\tan(\theta/2)\exp(-i\phi)$.

The representation operator $T^{j}(g)=\exp\left(-i\phi s_{1}^{(d)}\right)
\exp\left(-i\theta s_{2}^{(d)}\right)\exp\left(-i\psi s_{3}^{(d)}\right)$, $d=2j+1$, of the group $\mathrm{SU(2)}$ in the space $\mathcal{R}^{(d)}$, admits the decomposition
\begin{align}
	&  T^{j}(g)=D^{(d)}(\zeta)T^{j}(\psi),\ \ \zeta\in\mathbb{C}, 
	\quad 0\leq\psi<2\pi\ ,\nonumber\\
	&  D^{(d)}(\zeta)=\exp\left(  \zeta s_{+}^{(d)}\right)  \exp\left[  \ln\left(
	1+\left\vert \zeta\right\vert ^{2}\right)  s_{3}^{(d)}\right]  \exp\left(
	-\zeta^{\ast}s_{-}^{(d)}\right)  \ ,\nonumber\\
	&  T^{j}(\psi)=\exp\left(  -i\psi s_{3}^{(d)}\right)  ,\ \ s_{\pm}^{(d)}%
	=s_{1}^{(d)}\pm is_{2}^{(d)}\ . 
	\label{3.3.3}
\end{align}
Consider the action of the operators $T^{j}(g)$ on the fixed vector
$\left\vert j,-j\right\rangle^{(d)}$,
\begin{equation}
	T^{j}(g)\left\vert j,-j\right\rangle ^{(d)}=\exp\left[  i\left(  d-1\right)
	\psi/2\right]  D^{(d)}(\zeta)\left\vert j,-j\right\rangle ^{(d)}\ .
	\label{3.3.4}
\end{equation}
Vectors (\ref{3.3.4}) for different $\psi$, parameterizing the subgroup
$\mathrm{U(1)}$, differ only in a phase and define the same state. Let us set
$\psi=0$ in Eq. (\ref{3.3.4}). Then instantaneous \textrm{PSCS} are specified
by a point of the homogeneous space $X$ and have the form:
\begin{equation}
	\left\vert \zeta\right\rangle ^{(d)}=D^{(d)}(\zeta)\left\vert
	j,-j\right\rangle ^{(d)}\ . 
	\label{2.12b}
\end{equation}
The instantaneous \textrm{PSCS} are parameterized by a complex number $\xi\in\mathbb{C}$
and form a complete set of states,
\begin{equation}
	\int\left\vert \zeta\right\rangle ^{(d)\ (d)}\left\langle \zeta\right\vert
	\ d\mu_{j}\left(  \zeta\right)  =1,\ \ d\mu_{j}\left(  \zeta\right)
	=\frac{2j+1}{\pi}\frac{d\operatorname{Re}\zeta\ d\operatorname{Im}\zeta
	}{\left(  1+\left\vert \zeta\right\vert ^{2}\right)  ^{2}}\ . 
	\label{2.13b}
\end{equation}
Note that the uncertainty relation is minimized for for instantaneous
\textrm{PSCS},
\[
	\left\langle \tilde{s}_{1}^{(d)2}\right\rangle \left\langle \tilde{s}%
	_{2}^{(d)2}\right\rangle \geq\frac{1}{4}\left\langle \tilde{s}_{3}%
	^{(d)}\right\rangle ^{2},\ \ \tilde{s}_{k}^{(d)}=D^{(d)}(\zeta)s_{k}%
	^{(d)}D^{(d)^{-1}}(\zeta)\ .
\]

The states $\left\vert \zeta(t)\right\rangle ^{(d)}$ are called time dependent
\textrm{PSCS}. The function $\zeta(t)$ satisfy a special equation that
includes the external field in which the system is placed.

It is shown in Appendix \textrm{B} that the action of the operator $\left[
D^{(d)}(\zeta(t))\right]^{-1}$ for $\zeta(t)=\tilde{Z}_{1}(t)/\tilde{Z}_{2}(t)$ on the state $\left\vert Z,t\right\rangle^{(d)}$ yields a vector proportional to the vector $\left\vert j,-j\right\rangle^{(d)}=\overline{\left\vert d\right\rangle}$,
\begin{align}
	&  \ \left[  D^{(d)}(\zeta(t))\right]  ^{-1}\left\vert Z,t\right\rangle
	^{(d)}=\sum_{a,b=1}^{d}\overline{\langle a|}\left[  D^{(d)}(\zeta(t))\right]
	^{-1}\overline{|b\rangle}\ \overline{\langle b}\left\vert Z,t\right\rangle
	\ \overline{|a\rangle}\nonumber\\
	&  =\frac{\left\vert Z\right\vert ^{d-1}}{\sqrt{\left(  d-1\right)  !}%
	}e^{-\left\vert Z\right\vert ^{2}/2}\left(  \frac{\tilde{Z}_{2}(t)}{\tilde
		{Z}_{2}^{\ast}(t)}\right)  ^{\frac{d-1}{2}}\left\vert j,-j\right\rangle
	^{(d)},\ \ \zeta(t)=\frac{\tilde{Z}_{1}\left(  t\right)  }{\tilde{Z}%
		_{2}\left(  t\right)  }\ . 
	\label{2.17b}
\end{align}
Eq. (\ref{2.17b}) allows us to find a relation between the \textrm{AMCS} and
time-dependent \textrm{PSCS},
\begin{equation}
	\left\vert Z,t\right\rangle ^{(d)}=\frac{\left\vert Z\right\vert
		^{d-1}e^{-\left\vert Z\right\vert ^{2}/2}}{\sqrt{\left(  d-1\right)  !}%
	}e^{i(d-1)\arg\tilde{Z}_{2}(t)}\left\vert \zeta(t)\right\rangle ^{(d)}%
	,\ \ \zeta(t)=\frac{\tilde{Z}_{1}\left(  t\right)  }{\tilde{Z}_{2}\left(
		t\right)  }\ . 
	\label{2.18b}
\end{equation}

Note that the average values of time-independent physical quantities $L$
are proportional to their average values with respect to time-independent
\textrm{PSCS} $\left\vert \zeta\right\rangle^{(d)}$,
\begin{equation}
	^{(d)}\langle Z,t|\hat{L}|Z,t\rangle^{(d)}=\frac{\left\vert Z\right\vert
		^{2(d-1)}}{\left(  d-1\right)  !}e^{-\left\vert Z\right\vert ^{2}}%
	\ ^{(d)}\langle\zeta(t)|\hat{L}|\zeta(t)\rangle^{(d)}\ . 
	\label{2.18c}
\end{equation}

Constructed \textrm{AMCS} represent a non-trivial generalization of
\textrm{PSCS}. The states $\left\vert \zeta(t)\right\rangle ^{(d)}$ are set by
one time-dependent parameter $\zeta(t)$, whereas the states $\left\vert
Z,t\right\rangle ^{(d)}$ by two parameters $\tilde{Z}_{1}(t)$ and $\tilde{Z}_{2}(t)$. The only one subset of \textrm{AMCS} can be related to \textrm{PSCS}. This reflects the fact that the set of possible \textrm{AMCS} is wider than the set of \textrm{PSCS}, this is feature of our method of constructing \textrm{CS} for quadratic systems noted
in Ref. \cite{BagGiP2015}. As was already said exact solutions for
\textrm{AMCS} can be easily constructed on the base of exact solutions of
\textrm{SE} in $2$-dimensions that were found in Ref. \cite{BagGiBL05}. The
question of constructing exact solutions for \textrm{PSCS} remains open since
it is connected with finding external fields that allow exact solutions for
the parameter $\zeta(t)$ in the Perelomov approach.

\section{CS of spin 1/2 in a constant magnetic field\label{S4.1}}

As an example, let us consider the particular case of the moment $j=1/2$ in a
constant magnetic field\footnote{We not that coherent spin states in constant
	magnetic field were considered in frame work of different approaches in a
	number of works, see, e.g. Refs. \cite{CSmag,CSmag2,CSmag4,Perel86,Berge92}.}
$\mathbf{B}=(0,0,B)$ directed along the axis $z$. Then the corresponding
vector $\mathbf{F}$ in Eq. (\ref{1.5}) reads:
\begin{equation}
	\mathbf{F}=\left(  0,0,F\right)  ,\text{\ }F=2\omega_{0},\ \omega_{0}%
	=\frac{eB}{2m_{e}}. 
	\label{2.33}
\end{equation}

According to Eq. (\ref{1.24}), the corresponding equation for finding matrices
$w$ has the form:
\begin{align}
	&  i\dot{W}_{\alpha}\left(  t\right)  =\left(  \boldsymbol{\sigma\Omega
	}\right)  W_{\alpha}\left(  t\right)  ,\ \ W_{\alpha}\left(  t\right)
	=\left(
	\begin{array}
		[c]{c}%
		u_{\alpha1}\left(  t\right)  \\
		u_{\alpha2}\left(  t\right)
	\end{array}
	\right)  ,\ \boldsymbol{\Omega}=-\left(  0,0,\omega_{0}\right)  \nonumber\\
	&  \ \Longrightarrow i\dot{W}_{\alpha}\left(  t\right)  =-\omega_{0}\sigma
	_{3}W_{\alpha}\left(  t\right)  \ .
	\label{2.34}
\end{align}
Its general solution reads:
\begin{equation}
	W_{\alpha}\left(  t\right)  =\exp\left(  i\omega_{0}\sigma_{3}t\right)
	W_{\alpha}\left(  0\right)  \ .
	\label{2.35}
\end{equation}
Representing the spinors $W_{\alpha}(0)$ as:
\begin{equation}
	W_{\alpha}\left(  0\right)  =\left(  c_{\alpha}\right)  _{1}\left(
	\begin{array}
		[c]{c}
		1\\
		0
	\end{array}
	\right)  +\left(  c_{\alpha}\right)  _{2}\left(
	\begin{array}
		[c]{c}
		0\\
		1
	\end{array}
	\right)  \ ,
	\label{2.36}
\end{equation}
we obtain:
\begin{equation}
	W_{\alpha}\left(  t\right)  =\left(  c_{\alpha}\right)  _{1}\exp\left(
	i\omega_{0}t\right)  \left(
	\begin{array}
		[c]{c}
		1\\
		0
	\end{array}
	\right)  +\left(  c_{\alpha}\right)  _{2}\exp\left(  -i\omega_{0}t\right)
	\left(
	\begin{array}
		[c]{c}
		0\\
		1
	\end{array}
	\right)  \ .
	\label{2.37}
\end{equation}
Taking into account the initial conditions (\ref{1.26a}) for the matrices $w$, 
i.e., $w_{\alpha\beta}(0) = \delta_{\alpha\beta}$, we obtain:
$(c_{\alpha})_{\beta}=\delta_{\alpha\beta}$. Then:
\begin{equation}
	W_{\alpha}\left(  t\right)  =\left(
	\begin{array}
		[c]{l}%
		\delta_{\alpha1}\exp\left(  i\omega_{0}t\right)  \\
		\delta_{\alpha2}\exp\left(  -i\omega_{0}t\right)
	\end{array}
	\right)  \ ,
	\label{2.38}
\end{equation}
which implies:
\begin{align}
	&  w_{\alpha1}\left(  t\right)  =\delta_{\alpha1}\exp\left(  i\omega
	_{0}t\right)  ,\ w_{\alpha2}\left(  t\right)  =\delta_{\alpha2}\exp\left(
	-i\omega_{0}t\right)  \nonumber\\
	&  \ \Longrightarrow w_{\alpha\beta}\left(  t\right)  =\left(
	\begin{array}
		[c]{cc}%
		\exp\left(  i\omega_{0}t\right)   & 0\\
		0 & \exp\left(  -i\omega_{0}t\right)
	\end{array}
	\right)  \ .
	\label{2.39}
\end{align}
Thus, the \textrm{AMCS} of spin $1/2$ in a constant magnetic field has the
form:
\begin{align}
	\left\vert Z,t\right\rangle ^{(2)} &  =\exp\left\{  -\frac{|Z_{1}|^{2}%
		+|Z_{2}|^{2}}{2}\right\}  \left[  Z_{1}\exp\left(  -i\omega_{0}t\right)
	\overline{\left\vert 1\right\rangle }+Z_{2}\exp\left(  i\omega_{0}t\right)
	\overline{\left\vert 2\right\rangle }\right]  \ ,\label{2.40}\\
	\left\vert \mathbf{Z,t}\right\rangle ^{(2)} &  =\exp\left\{  -\frac
	{|Z_{1}|^{2}+|Z_{2}|^{2}}{2}\right\}  \left(
	\begin{array}
		[c]{c}%
		Z_{1}\exp\left(  -i\omega_{0}t\right)  \\
		Z_{2}\exp\left(  i\omega_{0}t\right)
	\end{array}
	\right)  \ .\nonumber
\end{align}
Then Eq. (\ref{2.18b}) takes the form:
\begin{align*}
	&  \left\vert Z,t\right\rangle ^{(2)}=\exp\left\{  -\frac{|Z_{1}|^{2}%
		+|Z_{2}|^{2}}{2}\right\}  \sqrt{\frac{|Z_{1}|^{2}+|Z_{2}|^{2}}{2}}\exp\left[
	i\left(  \arg Z_{2}+\omega_{0}t\right)  \right]  \left\vert \zeta
	(t)\right\rangle ^{(2)}\ ,\\
	&  \zeta(t)=\frac{Z_{1}}{Z_{2}}\exp\left(  -2i\omega_{0}t\right)  \ ,
\end{align*}
where the \textrm{PSCS} $\left\vert \zeta\right\rangle ^{(2)}$ are:
\begin{equation}
	\left\vert \zeta\right\rangle ^{(2)}=\sqrt{\frac{2}{1+\left\vert
			\zeta\right\vert ^{2}}}\ \left[  \zeta\overline{\left\vert 1\right\rangle
	}+\overline{\left\vert 2\right\rangle }\right]  \ .
	\label{1.51}
\end{equation}

We note that in Ref. \cite{Arecchi 1972} $\mathrm{CS}$ for a $2$-level system are
constructed as $\mathrm{CS}$ of angular momentum, sometimes they are called
atomic $\mathrm{CS}$. The evolution of these states is described by a point on
a unit two-dimensional sphere (on the Bloch sphere). Atomic $\mathrm{CS}$ in
angular parametrization are called Bloch $\mathrm{CS}$ and coincide with the
\textrm{PSCS}.

\section{Summary\label{S4}}

A way of introducing $\textrm{CS}$ of finite-level systems with a given
angular momentum in an external electromagnetic field is proposed. First we
consider an analogue of the spin equation in $2$-dimensional space in an
infinite-dimensional Hilbert space $\mathcal{R}^{(\infty)}$.
Such a space is constructed with the help of two kinds of Schwinger creation
and annihilation operators. The introduced equation is an equation for the
angular momentum in $\mathcal{R}^{(\infty)}$. One other side we
treat it as a generating spin equation ($\mathrm{GSE}$). By projecting the
$\mathrm{GSE}$ on $d$-dimensional subspaces, we obtain an analog of
$\mathrm{SE}$ for $d$-dimensional systems in an external electromagnetic
field. The Hamiltonian of $\mathrm{GSE}$ turns out to be quadratic in the
Schwinger creation and annihilation operators. This fact allows us to use a
modification of Malkin-Manko integral of motion method developed in our
earlier work for constructing the corresponding time-dependent generalized
$\mathrm{CS}$ for the quadratic system in $\mathcal{R}^{(\infty)}$. 
Projections of the later $\mathrm{CS}$ on finite-dimensional subspaces
describe $d$-dimensional system with a given angular moment, we call
\textrm{AMCS} of finite-level systems. The new \textrm{AMCS} for
$d$-dimensional systems in an external electromagnetic field admit the dynamic
symmetry group $SU(2)$. We note that the projection operation can
always be performed if the external field is sufficiently smooth. The set
\textrm{AMCS} is complete in $\mathcal{R}^{(d)}$. The \textrm{AMCS} $\left\vert Z,t\right\rangle ^{(d)}$ have a clear physical meaning, they obey the Schr\"odinger for a $d$-dimensional system with a given angular moment $j=(d-1)/2$ in an external electromagnetic field. Their possible exact solutions are constructed via exact solutions of the $\mathrm{SE}$ in $2$-dimensional space. The latter solutions can be found analytically and are completely described in our earlier works. The only one
subset of \textrm{AMCS} can be related to \textrm{PSCS}. This reflects the
fact that the set of possible \textrm{AMCS} is wider than the set of
\textrm{PSCS}, this is feature of our method of constructing \textrm{CS}
for quadratic systems noted in Ref. \cite{BagGiP2015}. The question of
constructing exact solutions for \textrm{PSCS} remains open since it is
connected with finding external fields that allow exact solutions for the
parameter $\zeta(t)$ in the Perelomov approach.

The results obtained in the present work may be interesting in the context of
revealing a connection between two different methods of constructing
$\mathrm{CS}$. According to Perelomov, in the case of the $SU(2)$ group, spin $\mathrm{CS}$ are introduced by the group-theoretic method,
namely, by the action of representation operators of the group on a fixed
vector. In our construction $\mathrm{CS}$ of angular momentum $j$ are given by
projections of a $\mathrm{CS}$ constructed, in fact, by Malkin-Manko method,
in an infinite-dimensional Fock space onto its $(2j+1)$-dimensional subspace.
The above mentioned relation between \textrm{PSCS} and \textrm{AMCS}
establishes, in a sense, a relation between the Perelomov and the Malkin-Manko
methods. \textrm{AMCS} states in a constant magnetic field are constructed.
Some of them coincide with the Bloch $\mathrm{CS}$.

\section*{Acknowledgments}

D.M.G. thanks CNPq for permanent support.

\section*{Appendix A\label{AppB}}

Let us demonstrate that the subspace $\mathcal{R}^{(d)}$ is
invariant under the action of the angular momentum operators (\ref{1.8}) and
find their matrix elements in $\mathcal{R}^{(d)}$.

It follows from Eq. (\ref{1.13ee}) that
\begin{equation}
	\hat{a}_{\alpha}^{\dagger}\hat{a}_{\beta}\left\vert \Psi\right\rangle
	^{\left(  d\right)  }=\sum_{a=1}^{d}\psi_{a}^{(d)}\ \hat{a}_{\alpha}^{\dagger
	}\hat{a}_{\beta}\overline{\left\vert a\right\rangle }\ . 
	\label{B.1}
\end{equation}
The operators $\hat{a}_{\alpha}^{\dagger}\hat{a}_{\beta}$ act on the
ket-vectors (\ref{1.13bb}) in the following way:
\begin{align}
	&  \hat{a}_{\alpha}^{\dagger}\hat{a}_{\beta}\overline{\left\vert
		a\right\rangle }=\hat{a}_{\alpha}^{\dagger}\hat{a}_{\beta}\left\vert
	d-a,a-1\right\rangle =\delta_{\alpha,\beta}\left[  \left(  d-a\right)
	\delta_{\alpha,1}+\left(  a-1\right)  \delta_{\alpha,2}\right]  \overline
	{\left\vert a\right\rangle }\nonumber\\
	&  +\delta_{\alpha,2}\delta_{\beta,1}\sqrt{a(d-a)}\overline{\left\vert
		a+1\right\rangle }+\delta_{\alpha,1}\delta_{\beta,2}\sqrt{(a-1)(d-a+1)}%
	\overline{\left\vert a-1\right\rangle }\ . 
	\label{B.2}
\end{align}
As it follows from Eq. (\ref{B.2}) resulting vectors remain in the same space,
\begin{align}
	&  \ \hat{a}_{\alpha}^{\dagger}\hat{a}_{\beta}\left\vert \Psi\right\rangle
	^{\left(  d\right)  }=\sum_{a=1}^{d}\left\{  \delta_{\alpha,\beta}\left[
	\left(  d-a\right)  \delta_{\alpha,1}+\left(  a-1\right)  \delta_{\alpha
		,2}\right]  \psi_{a}^{(d)}\right. \nonumber\\
	&  +\left.  \delta_{\alpha,2}\delta_{\beta,1}\sqrt{(a-1)(d-a+1)}\psi
	_{a-1}^{(d)}+\delta_{\alpha,1}\delta_{\beta,2}\sqrt{a(d-a)}\psi_{a+1}%
	^{(d)}\right\}  \overline{\left\vert a\right\rangle }\ . 
	\label{B.3}
\end{align}
From here follows the invariance of operators (\ref{1.8}).

The matrix elements of the operators $\hat{a}_{\alpha}^{\dagger}\hat{a}_{\beta}$ are:
\begin{align}
	&  \overline{\langle a|}\hat{a}_{\alpha}^{\dagger}\hat{a}_{\beta}%
	\overline{|b\rangle}=\left[  \left(  d-a\right)  \delta_{\alpha,1}+\left(
	a-1\right)  \delta_{\alpha,2}\right]  \delta_{a,b}\delta_{\alpha,\beta
	}\nonumber\\
	&  +\sqrt{a(d-a)}\delta_{a+1,b}\delta_{\alpha,2}\delta_{\beta,1}%
	+\sqrt{(a-1)(d-a+1)}\delta_{a-1,b}\delta_{\alpha,1}\delta_{\beta,2}\ .
	\label{B.4}
\end{align}
From Eq. (\ref{B.4}) it is easy to obtain matrix elements of angular momentum
operators (\ref{1.8}),
\begin{align}
	\left(  \hat{S}_{i}\right)  _{ab}  &  =\overline{\langle a|}\hat{S}%
	_{i}\overline{|b\rangle}=(\sigma_{i}^{\alpha\beta}/2)\overline{\langle a|}%
	\hat{a}_{\alpha}^{\dagger}\hat{a}_{\beta}\overline{|b\rangle}\label{B.5}\\
	\left(  \hat{S}_{1}\right)  _{ab}  &  =\frac{1}{2}\left[  \sqrt{a(d-a)}%
	\delta_{a+1,b}+\sqrt{b(d-b)}\delta_{a,b+1}\right]  \ ,\nonumber\\
	\left(  \hat{S}_{2}\right)  _{ab}  &  =\frac{i}{2}\left[  \sqrt{b(d-b)}%
	\delta_{a,b+1}-\sqrt{a(d-a)}\delta_{a+1,b}\right]  \ ,\nonumber\\
	\left(  \hat{S}_{3}\right)  _{ab}  &  =\frac{1}{2}\left(  d-2a+1\right)
	\delta_{a,b}\ , 
	\label{B.6}
\end{align}
such that
\begin{equation}
	\left(  \mathbf{\hat{S}}^{2}\right)  _{ab}=\left(  \hat{S}_{1}^{2}\right)
	_{ab}+\left(  \hat{S}_{2}^{2}\right)  _{ab}+\left(  \hat{S}_{3}^{2}\right)
	_{ab}=j(j+1)\delta_{a,b},\ \ j=\frac{d-1}{2}\ . 
	\label{B.7}
\end{equation}
The following matrix elements are used in our constructions:
\begin{align}
	\overline{\langle a|}\exp\left(  \zeta\hat{S}_{+}\right)  \overline
	{|b\rangle}  &  =\sum_{k=0}^{d-1}\sqrt{\frac{(a+k-1)!(d-a)!}{(d-k-a)!(a-1)!}%
	}\frac{\zeta^{k}}{k!}\delta_{a+k,b},\ \ \overline{\langle a|}\exp\left(
	\hat{S}_{3}\ln\zeta\right)  \overline{|b\rangle}=\zeta^{\frac{d-2a+1}{2}%
	}\delta_{a,b}\ ,\nonumber\\
	\overline{\langle a|}\exp\left(  \zeta\hat{S}_{-}\right)  \overline
	{|b\rangle}  &  =\sum_{l=0}^{d-1}\sqrt{\frac{(b+l-1)!(d-b)!}{(d-l-b)!(b-1)!}%
	}\frac{\zeta^{k}}{k!}\delta_{a,b+l},\ \ \zeta\in\mathbb{C}\ . 
	\label{B.8}
\end{align}

\section*{Appendix B\label{AppC}}

Let us demonstrate that the action of the operator (with $\zeta(t)=\tilde{Z}_{1}(t)/\tilde{Z}_{2}(t)$)
\begin{equation}
	\left[  D^{(d)}(\zeta(t))\right]  ^{-1}=\exp\left[  -\zeta^{\ast}%
	(t)s_{-}^{(d)}\right]  \exp\left[  \ln\left(  1+\left\vert \zeta(t)\right\vert
	^{2}\right)  s_{3}^{(d)}\right]  \exp\left[  \zeta(t)s_{+}^{(d)}\right]
	\label{C.1}
\end{equation}
on the state $\left\vert Z,t\right\rangle ^{(d)}$ (see Eq. (\ref{2.8}))
produces vector proportional to the vector 
$\left\vert j,-j\right\rangle^{(d)}=\overline{\left\vert d\right\rangle}$.

To this end we first study matrix elements of operator (\ref{C.1}):
\begin{align}
	&  \overline{\langle a|}\left[  D^{(d)}(\zeta)\right]  ^{-1}\overline
	{|b\rangle}\nonumber\\
	&=\sum_{c,e=1}^{d}\overline{\langle a|}\exp\left[  -\zeta^{\ast
	}(t)s_{-}^{(d)}\right]  \overline{|c\rangle}\overline{\langle c|}\exp\left[
	\ln\left(  1+\left\vert \zeta(t)\right\vert ^{2}\right)  s_{3}^{(d)}\right]
	\overline{|e\rangle}\overline{\langle e|}\exp\left[  \zeta(t)s_{+}%
	^{(d)}\right]  \overline{|b\rangle}\nonumber\\
	&  =\sum_{c,e=1}^{d}\overline{\langle a|}\exp\left[  -\zeta^{\ast}(t)\hat
	{S}_{-}\right]  \overline{|c\rangle}\overline{\langle c|}\exp\left[
	\ln\left(  1+\left\vert \zeta(t)\right\vert ^{2}\right)  \hat{S}_{3}\right]
	\overline{|e\rangle}\overline{\langle e|}\exp\left[  \zeta(t)\hat{S}%
	_{+}\right]  \overline{|b\rangle}\ . 
	\label{C.1b}
\end{align}
With the help of Eq. (\ref{B.8}), we obtain:
\begin{align}
	&  \overline{\langle a|}\left[  D^{(d)}(\zeta)\right]  ^{-1}\overline
	{|b\rangle}=\left(  1+\left\vert \zeta\right\vert ^{2}\right)  ^{-(d+1)/2}%
	\sum_{c=1}^{d}\frac{(d-c)!}{(c-1)!}\sqrt{\frac{(a-1)!(b-1)!}{(d-a)!(d-b)!}%
	}\nonumber\\
	&  \times\sum_{k,l=0}^{d-1}\left(  -\frac{1+\left\vert \zeta\right\vert ^{2}%
	}{\left\vert \zeta\right\vert ^{2}}\right)  ^{c}\frac{(\zeta^{\ast}%
		)^{a}(-\zeta)^{b}}{(a-c)!(b-c)!}\delta_{a-c,l}\delta_{b-c,k}\ . 
	\label{C.2}
\end{align}

Then we consider the following equation:
\begin{align}
	&  \overline{\langle a}|\left[  D^{(d)}(\zeta(t))\right]  ^{-1}\left\vert
	Z,t\right\rangle ^{(d)}=\sum_{b=1}^{d}\overline{\langle a|}\left[
	D^{(d)}(\zeta(t))\right]  ^{-1}\overline{|b\rangle}\ \overline{\langle
		b}\left\vert Z,t\right\rangle ^{(d)}\nonumber\\
	&  =\frac{\left[  \zeta^{\ast}(t)\right]  ^{a}e^{-\left\vert Z\right\vert
			^{2}/2}}{\left(  1+\left\vert \zeta(t)\right\vert ^{2}\right)  ^{(d+1)/2}%
	}\frac{\tilde{Z}_{1}^{d}(t)}{\tilde{Z}_{2}(t)}\sqrt{\frac{(a-1)!}{(d-a)!}%
	}\nonumber\\
	&  \times\sum_{c=1}^{d}\frac{(d-c)!}{(a-c)!(c-1)!}\left(  -\frac{1+\left\vert
		\zeta(t)\right\vert ^{2}}{\left\vert \zeta(t)\right\vert ^{2}}\right)
	^{c}\sum_{b=1}^{d}\ \frac{\left(  -1\right)  ^{b}}{(b-c)!(d-b)!}\sum
	_{k,l=0}^{d-1}\delta_{a-c,l}\delta_{b-c,k}\ . 
	\label{C.3}
\end{align}
Taking into account that
\begin{equation}
	\sum_{k,l=0}^{d-1}\delta_{a-c,l}\delta_{b-c,k}=\left\{
	\begin{array}
		[c]{l}%
		1,\ a\geq c,\ \ b\geq c\\
		0,\ \emph{otherwise}%
	\end{array}
	\right.  , 
	\label{C.4}
\end{equation}
we obtain
\begin{equation}
	\sum_{b=1}^{d}\ \frac{\left(  -1\right)  ^{b}}{(b-c)!(d-b)!}\sum_{k,l=0}%
	^{d-1}\delta_{a-c,l}\delta_{b-c,k}=(-1)^{d}\delta_{a,d}\delta_{c,d}\ .
	\label{C.5}
\end{equation}
Substituting Eq. (\ref{C.5}) into Eq. (\ref{C.3}), we find:
\begin{equation}
	\overline{\langle a}|\left[  D^{(d)}(\zeta(t))\right]  ^{-1}\left\vert
	Z,t\right\rangle ^{(d)}=\frac{\left\vert Z\right\vert ^{d-1}}{\sqrt{(d-1)!}%
	}e^{-\left\vert Z\right\vert ^{2}/2}\left(  \frac{\tilde{Z}_{2}(t)}{\tilde
		{Z}_{2}^{\ast}(t)}\right)  ^{\frac{d-1}{2}}\delta_{a,d}\ . 
	\label{C.6}
\end{equation}
Finally, Eq., (\ref{C.6}) allows us to obtain the following result that is
used in our constructions:
\begin{align}
	&  \left[  D^{(d)}(\zeta(t))\right]  ^{-1}\left\vert Z,t\right\rangle
	^{(d)}=\sum_{a,b=1}^{d}\overline{\langle a|}\left[  D^{(d)}(\zeta(t))\right]
	^{-1}\overline{|b\rangle}\ \overline{\langle b}\left\vert Z,t\right\rangle
	\ \overline{|a\rangle}\ \nonumber\\
	&  =\frac{\left\vert Z\right\vert ^{d-1}e^{-\left\vert Z\right\vert ^{2}/2}%
	}{\sqrt{\left(  d-1\right)  !}}\left(  \frac{\tilde{Z}_{2}(t)}{\tilde{Z}%
		_{2}^{\ast}(t)}\right)  ^{\frac{d-1}{2}}\left\vert j,-j\right\rangle
	^{(d)}\ ,\ \ \zeta(t)=\frac{\tilde{Z}_{1}\left(  t\right)  }{\tilde{Z}%
		_{2}\left(  t\right)  }\ . 
	\label{C.7}
\end{align}

\end{document}